\documentclass[12pt,float,cite]{iopart}
\usepackage{graphicx}
\begin{document}
%
%
\title[The phase transition of the diffusive pair contact process revisited]
      {The phase transition of the diffusive pair contact process revisited}
\author{Haye Hinrichsen}

\address{Fakult\"at f\"ur Physik und Astronomie, 
         Universit\"at W\"urzburg, D-97074 W\"urzburg, Germany}

\begin{abstract}
The restricted diffusive pair contact process $2A\to3A, \, 2A\to\emptyset$ (PCPD) and the classification of its critical behavior continues to be a challenging open problem of non-equilibrium statistical mechanics. Recently Kockelkoren and Chat{\'e} [Phys. Rev. Lett. {\bf 90}, 125701 (2003)] suggested that the PCPD in one spatial dimension represents a genuine universality class of non-equilibrium phase transitions which differs from previously known classes. To this end they introduced an efficient lattice model in which the number of particles per site is unrestricted. In numerical simulations this model displayed clean power laws, indicating ordinary critical behavior associated with certain non-trivial critical exponents. In the present work, however, we arrive at a different conclusion. Increasing the numerical effort, we find a slow drift of the effective exponents which is of the same type as observed in previously studied fermionic realizations. Analyzing this drift we discuss the possibility that the asymptotic critical behavior of the PCPD may be governed by an ordinary directed percolation fixed point.
\end{abstract}

\pacs{05.50.+q, 05.70.Ln, 64.60.Ht}
\maketitle
\def\xvec{{\mathbf{x}}}
\def\kvec{{\mathbf{k}}}
\def\text#1{\mbox{#1}}
\parskip 1mm 
%
\section{Introduction}
\label{intro}
%
In the attempt to classify non-equilibrium phase transitions from fluctuating phases into absorbing states~\cite{MarroDickman,Hinrichsen00a,OdorReview}, the so-called pair contact process with diffusion (PCPD) continues to attract considerable attention (for a recent review see Ref.~\cite{HenkelHinrichsen04}). The PCPD is a reaction-diffusion process that describes a single species of diffusing particles subjected to binary reactions of the form
\begin{equation}
\label{reactions}
2A \to 3A \,, \qquad 2A \to \emptyset
\end{equation}
combined with a mechanism which prevents the particle density from diverging. In one spatial dimension this process exhibits an unconventional type of critical behavior, raising the question whether the PCPD represents a novel  universality class of non-equilibrium phase transitions~\cite{Hinrichsen00b}. Numerical studies of so-called \textit{fermionic} lattice models, where each site can be occupied by at most one particle, displayed a slow drift of the effective critical exponents and so far turned out to be inconclusive. More recently, however, Kockelkoren and Chat{\'e} introduced a particularly efficient one-dimensional lattice model, in which the number of particles per site is unrestricted~\cite{KockelkorenChate}. In contrast to previously studied models, where the reactions involve three neighboring sites, the binary reactions~(\ref{reactions}) are carried out at \textit{single} sites, combined with  ordinary nearest-neighbor diffusion. Unlike fermionic realizations of the PCPD, the model introduced by Kockelkoren and Chat{\'e} was found to exhibit clean power laws over several decades. The estimated values of the corresponding critical exponents~\cite{KockelkorenChate}
\begin{equation}
\label{ChateExponents}
\delta = \frac{\beta}{\nu_\parallel}=0.200(5), \quad
z = \frac{\nu_\perp}{\nu_\parallel}=1.70(5), \quad
\beta = 0.37(2).
\end{equation}
led the authors to the conclusion that the PCPD represents a novel genuine universality class of non-equilibrium phase transitions which differs from all previously known classes. Likewise, they investigated various other reaction-diffusion rules. In particular they found that the so-called triplet process $3A\to 4A, \, 3A\to\emptyset$ introduced in~\cite{TripletProcess} shows yet another novel type of critical behavior. Combining all these results, Kockelkoren and Chat{\'e} presented a generic classification table for absorbing phase transitions that comprises four non-trivial classes, namely, directed percolation (DP), the parity-conserving (or voter) universality class, the PCPD, and the triplet process. 

The results of the present paper indicate that this classification scheme may be premature and needs to be reexamined. Highly optimized large-scale simulations of the model introduced in Ref.~\cite{KockelkorenChate} reveal that Kockelkoren and Chat{\'e}, apparently attempting to obtain straight lines in a double-logarithmic plot, systematically underestimated the critical threshold. Increasing the numerical effort it turns out that their model does not show clean power laws, instead it is plagued by the very same type strong deviations as observed in previously studied fermionic models. The effective critical exponents are found to display a slow drift that probably extends beyond the numerically accessible range, most likely approaching an ordinary directed percolation fixed point.

\section{Definition of the model}
\label{definition}
%
%
The class of models introduced by Kockelkoren and Chat{\'e} is best explained in the case of the PCPD in one spatial dimension (rule {\tt pp12} in their nomenclature). Consider a one-dimensional lattice of $L$ sites with periodic boundary conditions. Each site $i$ is associated with an integer number $n_i>0$ that represents the local number of particles. The dynamics is controlled by a single parameter $p\in [0,1]$ and evolves in two synchronized substeps as follows:
\begin{itemize}
\item[$\bullet$] \textbf{Diffusion:}
At first each particle diffuses independently with equal probability to one of the nearest neighbors.
\item[$\bullet$] \textbf{On-site reactions:}
In the second substep all sites $i$, which are occupied by at least two particles, $n_i\geq 2$, are updated individually. To this end the particles at a given site are grouped into $m_i=\lfloor n_i/2\rfloor$ pairs, where $\lfloor\cdot\rfloor$ denotes truncation to the largest integral value not greater than the argument. Each of these pairs independently produces one offspring with probability $p^{m_i}$, increasing $n_i$ by $1$, or annihilates otherwise, thus decreasing $n_i$ by $2$.
\end{itemize}
Obviously, when $p$ is small, annihilation dominates, leading to an algebraic decay of the density $\rho(t) \sim t^{-1/2}$ which characterizes the absorbing phase of the PCPD. On the other hand, if $p$ is sufficiently large, particle production dominates and the system approaches a stationary active phase. Note that for $p<1$ the generation of new particles is \textit{exponentially} suppressed as $n_i$ increases. This exponential cutoff, which is the counterpart of the fermionic constraint ($n_i\leq 1$) in previously studied PCPD models, prevents the particle density from diverging in the active phase. 

The update algorithm can be optimized as follows. Because of the exponential cutoff particle numbers greater than $n_i=16$ are extremely rare so that we can safely restrict their range to $0 \leq n_i < 32$. To speed up the on-site reactions we define for each $n_i$ a list of probabilities for the possible outcomes which are then selected by using a single random number. Furthermore, instead of implementing the lattice as a static array, the actual configuration is stored as a dynamically generated list of the coordinates of occupied sites. These two improvements, combined with some general technical optimizations, accelerate the update algorithm significantly. 

\section{Numerical results}
\label{numerical}
%
%
Let us consider simulations starting with a homogeneous initial state, where each site is occupied by a pair of particles. Kockelkoren and Chat{\'e} studied this case by performing a \textit{single} run on a large system with $L=2^{22} \approx 4.2\times 10^7$ sites simulating up to $t=2\times 10^6$ Monte Carlo updates (see Fig. 1 of Ref.~\cite{KockelkorenChate}). Plotting various order parameters in a double-logarithmic representation and seeking for asymptotically straight lines they estimated the critical point $p_c=0.795410(5)$. In the improved simulations presented here the temporal range is extended by almost one decade up to $t=1 \times 10^8$ Monte Carlo updates. Furthermore, using the same system size of $L=2^{22}$ sites as in Ref.~\cite{KockelkorenChate} we average over at least 40 independent realizations of randomness in order to reduce the statistical error. 
%
%
\begin{figure}
\centerline{\includegraphics[width=130mm,angle=270]{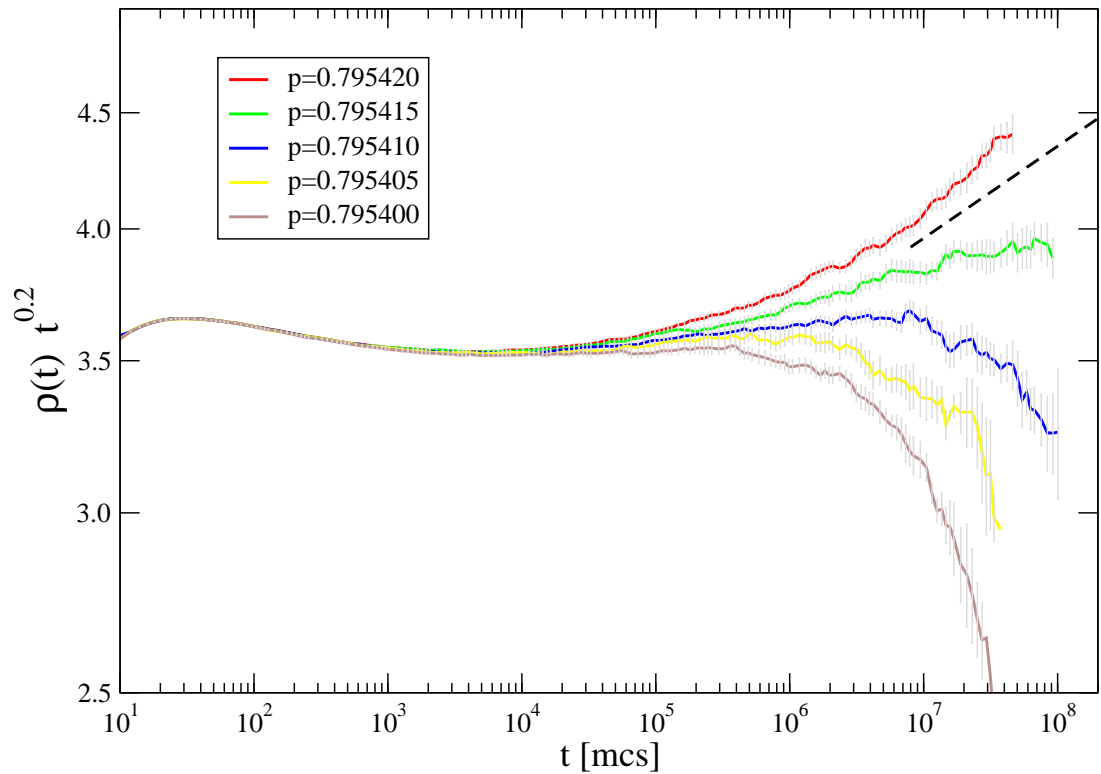}}
\caption{\label{FIGDECAY}
Density of particles starting with homogeneous initial conditions as a function of time multiplied by $t^{0.2}$ for various values of $p$. The error bars indicate the expected standard deviation averaged over the samples. Other order parameters as those studied in Ref.~\cite{KockelkorenChate} lead to similar results. The dashed line represents the slope that corresponds to the critical behavior of directed percolation.
}
\end{figure}
%
%
The results are shown in Fig.~\ref{FIGDECAY}. As can be seen, the curves for $p$ equal to $0.795405$, $0.795410$, and $0.795415$ bend downwards. Consequently the critical point $p_c=0.795410(5)$ given in~\cite{KockelkorenChate}, including its error margin, lies entirely in the inactive phase. It seems that Kockelkoren and Chat{\'e}, attempting to obtain an apparent power-law behavior, systematically underestimated the critical point.

Instead of clean power laws we observe a slow drift of the effective critical exponents which manifests itself as a weak curvature of the data in Fig.~\ref{FIGDECAY} that extends over at least four decades in time. Therefore the model introduced by Kockelkoren and Chat{\'e} is not a good-natured exception in the zoo of suggested PCPD models, rather it shows the same phenomenological properties as previously studied fermionic variants and is plagued by exactly the same type of slowly decaying corrections to scaling. At the present state of knowledge there is no reason to believe that the observed drift suddenly stops after $10^6$ or $10^7$ time steps, instead it seems to be more likely that the drift continues to vanish gradually over many decades beyond the numerically accessible temporal range. Therefore, it would be a mistake to search again for straight-looking tails of the curves and to specify updated exponents of a possible PCPD universality class. However, assuming that the drift eventually vanishes without changing sign, such estimates may serve as upper bounds of postulated asymptotic critical exponents. For example, analyzing the data shown in Fig.~\ref{FIGDECAY} one obtains $\delta<0.185$. Similar inequalities are obtained by finite-size and off-critical simulations, as summarized in Table~\ref{Tab1}.

\begin{table}
\begin{center}
\begin{tabular}{|c||c|c|c|c|}
\hline  & $p_c$ & $\delta$ & $z$ & $\beta$ \\ \hline
\hline Ref.~\cite{KockelkorenChate} & $0.795410(5)$ & $0.200(5)$ & $1.70(5)$ & $0.37(2)$ \\ 
\hline present work & $0.795417(2)$ & $<0.185$ & $<1.65$ & $<0.34$ \\ 
\hline DP & - & $0.1595$ & $1.5807$ & $0.2765$ \\ 
\hline 
\end{tabular}
\caption{\label{Tab1}Estimates for the critical exponents compared with DP exponents.}
\end{center}
\end{table}

\section{Discussion}
\label{discussion}
%
%
The numerical results presented in this paper demonstrate that the PCPD model introduced in Ref.~\cite{KockelkorenChate}, in contrast to earlier claims, does not reach the scaling regime in the numerically accessible temporal range, instead it displays a slow drift of the effective critical exponents, just in the same way as in previously studied 'fermionic' variants of the model with hard-core exclusion. Because of this drift the critical exponents estimated by Kockelkoren and Chat{\'e} turn out to be incorrect and have to be replaced by the inequalities given in Table~\ref{Tab1}.

As in all PCPD models the observed drift is so slow that any attempt to extrapolate the effective exponents to $t \to \infty$ involves to some extent speculative expectations. Strictly speaking it is even not yet clear whether such a limit exists, questioning the very existence of a second-order phase transition. However, in my opinion the numerical results strongly support earlier suggestions~\cite{CA,Barkema} that the one-dimensional PCPD -- including its realization by Kockelkoren and Chat{\'e} -- may eventually cross over to an ordinary directed percolation transition. 

The hypothesis of a  DP transition in the PCPD is based on the following argument (cf. Ref.~\cite{CA} for details). In the numerically accessible range the PCPD is neither reaction- nor diffusion-limited, instead one observes a spatio-temporal coexistence of clustered reactions and freely diffusing solitary particles~\cite{cyclic}. These single particles diffuse over large distances in the 1+1-dimensional geometry with the dynamic exponent $z=2$ and their diffusion constant is well-defined and can be read off easily at any time. On the other hand, all simulations in 1+1 dimensions consistently confirm that the critical cluster spreads superdiffusively, i.e., $z<2$. The assumption of a scale-free situation therefore leads to a contradiction unless the diffusion constant flows to zero under renormalization group transformations as the critical point is approached, i.e., to a situation which is known to belong to the directed percolation universality class. 

In the light of the present numerical results the classification scheme introduced by Kockelkoren and Chat{\'e} 
needs to be reexamined carefully by various methods. In particular it would be interesting to verify their conclusions concerning the triplet process by similar large-scale simulations.

\vspace{5mm}
\noindent
\textbf{Acknowledgements}\\
I would like to thank A. Klein and A. Vetter for exclusive access to a new 80-node Linux cluster at the University of W{\"u}rzburg as well as the Leibniz Computing Center in Munich, where parts of the simulations were carried out.

\vspace{5mm}
\noindent{\bf References}\\


\begin{thebibliography}{99}                               

\bibitem{MarroDickman}  
Marro J and Dickman R,
\newblock {\em Nonequilibrium phase transitions in lattice models}.
\newblock Cambridge University Press, Cambridge, 1999.

\bibitem{Hinrichsen00a}
Hinrichsen H, 
\textit{Nonequilibrium critical phenomena and phase transitions into absorbing states}, 
2000 Adv. Phys. {\bf 49} 815.

\bibitem{OdorReview} 
{\'O}dor G, 
\textit{Universality classes in nonequilibrium lattice systems},
2004 Rev. Mod. Phys. {\bf 76} 663.

\bibitem{HenkelHinrichsen04}
Henkel M and Hinrichsen H,
\textit{The non-equilibrium phase transition of the pair-contact process with diffusion},
2004 J. Phys. A {\bf 37}, R117-R159.

\bibitem{Hinrichsen00b}
Hinrichsen H ,
\textit{Pair-contact process with diffusion: A new type of nonequilibrium critical behaviour?}, 
2001 Phys. Rev. E {\bf 63}  036102.

\bibitem{KockelkorenChate}
Kockelkoren J and Chat{\'e} H,
\textit{Absorbing phase transition of branching-annihilating random walks}, 
2003 Phys. Rev. Lett. {\bf 90} 125701.

\bibitem{TripletProcess}
Park K, Hinrichsen H, and Kim I-M, 
\textit{Phase transition in a triplet process},
2002  Phys. Rev. E {\bf 65}  R025101.

\bibitem{cyclic}
Hinrichsen H, 
\textit{Cyclically coupled spreading and pair-annihilation},
2001 Physica A {\bf 291} 275.

\bibitem{CA}
Hinrichsen H, 
\textit{Stochastic cellular automaton for the coagulation-fission process} $2A \to 3A$, $2A \to A$, 
2003 Physica A {\bf 320} 249.

\bibitem{Barkema}
Barkema~G T and Carlon E,
\textit{Universality in the pair-contact process with diffusion}, 
2003  Phys. Rev. E {\bf 68}  036113.

\bibitem{twodimensions}
{\'O}dor G, Marques~M C, and Santos~M A, 
\textit{Phase transition of a two-dimensional binary spreading model}, 
2002 Phys. Rev. E {\bf 65} 056113.


\end{thebibliography}
\end{document}